\documentclass[]{iopart}

\usepackage{graphicx}

\newcommand{\be}{\begin{equation}}
\newcommand{\ee}[1]{\label{#1} \end{equation}}
\newcommand{\ba}{\begin{eqnarray}}
\newcommand{\nl}{\nonumber \\}
\newcommand{\ea}[1]{\label{#1} \end{eqnarray}}

\begin{document}

\title[Pions and kaons]{Pions and kaons from stringy quark matter}

\author{T.~S.~Bir\'o and K.~\"Urm\"ossy}
\address{MTA KFKI Research Institute for Particle and Nuclear Physics \\ H-1525 Budapest, P.O.Box 49.}
\eads{\mailto{tsbiro@rmki.kfki.hu}, \mailto{ukaroly@freemail.hu}}


\begin{abstract}
 Different hadron transverse momentum spectra are calculated in a non-extensive statistical, 
 quark-coalescence model. We suggest that extreme relativistic kinematics may be
 responsible for power-law tails under quite general assumptions and a quark scaling
 could be recognized in the powers. For the low energy part also a gluonic string 
 contribution is needed to describe experimental data: its length distribution and
 fractality are fitted.
\end{abstract}


\section{Introduction}

The hadronization of quark matter in relativistic heavy ion collisions is a key
phenomenon for understanding the interplay between quark confinement and deconfinement,
between statistical and dynamical factors. The purely hadronic statistical model
\cite{SM1,SM2,SM3,SM4,SM5} was successful in describing several number ratios of different hadron species
and - by assuming a collective flow - some bulk properties of individual hadron spectra.
This model fits experimental results at several bombarding energies ranging from SIS to RHIC,
and finds that the main parameters, the temperature $T$ and the baryochemical potential, $\mu$
lies in a narrow range near to a line described by a constant energy per particle of $E/N=1$ GeV.
This value is six times larger than the apparent temperature of spectra after subtracting
transverse flow effects. This raises the question that how can be a quark matter, describable
by perturbative QCD and containing nearly massless quarks and massless gluons, transformed
into a hadronic matter with an average mass of $\overline{m}=750$ MeV (based on the simple
nonrelativistic ideal gas formula $E/N=m+3T/2$).

Recently we have suggested an approach to massless quark matter supplemented with
stringy interaction corrections\cite{BiroCleymans2008} giving a contribution to the
free energy density which is proportional to the (colored) number density on a fractional
power. It was shown that such an equation of state accommodates the $E/N=6T$ ratio at
zero pressure, at the edge of mechanical stability. This model assumes underlying
non-perturbative effects around and above the color deconfinement temperature, $T_c \approx 170$
MeV. Quark matter equation of state data from
lattice QCD support an interaction modification to the ideal massless gas picture
of $f_{{\rm int}} \propto n^{2/3}$. Since the basic concept is quite simple, we repeat
briefly this argumentation in the present paper.

On the other hand the quark coalescence picture proved to be applicable to the hadronization
problem in relativistic heavy ion collisions at RHIC energies in several respects.
The quark number scaling of asymmetric azimuthal elliptic flow factor\cite{v2a,v2b,v2c,pc1,pc2}, 
$v_2$, and the number ratios of heavy flavor, most prominently strange, hadrons to the light ones,
as it has been repeatedly reported in calculations using the ALCOR model\cite{ALCOR2008},
all underline the viability of this concept. We also wish to add a further observable,
namely the Tsallis parameter, $q-1$, regulating a cut power-law fit to transverse momentum
spectra, to this list. In this paper we demonstrate that this parameter also may follow
a quark number scaling (coalescence) law. 

In order to interpret this result in proper context, also some introduction is necessary
into recent developments on the interpretation of non-extensive thermodynamics.
We give a comprised summary of investigating the thermodynamical limit of abstract
composition rules\cite{BiroEPL2008} for two particle systems, which motivates the use of cut power-law
distributions of quark momenta in the framework of statistical physics models.
By doing so we connect predictions from dynamical parton models for $p_T$-spectra to
those of simpler statistical models smoothly\cite{Biro:2008km}.

\section{String distribution and equation of state in quark matter}

Assuming a pair interaction energy also in the quark matter,
which is proportional to the nearest neighbour distance,
one deals with an energy density modification of $n \sigma \langle \ell \rangle = An^{2/3}$.
The free energy density becomes
\be
 f(n,T) = f_{{\rm id}}(n,T) + An^{2/3}.
\ee{STRINGFREE}
Deriving the energy density and pressure via standard thermodynamical formulas one obtains
density dependent, stringy mean field terms, which - unlike in the bag model - do not compensate
each other exactly:
\be
 e = e_{{\rm id}} + \frac{3}{2} An^{2/3},  \qquad \qquad p = p_{{\rm id}} - \frac{1}{2} A n^{2/3}.
\ee{EOP}
At vanishing pressure the correction can be related to the ideal pressure by $p=0$ and
the energy per particle for massless constituents in a Boltzmann gas ($p_{{\rm id}}=nT$)
is given by
\be
 \frac{E}{N} = \left. \frac{e}{n} \right|_{p=0} =
  \left. \frac{e_{{\rm id}}+3p_{{\rm id}}}{n} \right|_{m=0} =
   \frac{6p_{{\rm id}}}{n} = 6T.
\ee{E6T}
We note that any interaction correction proportional to $n^{1-\gamma}$ in the free energy
leads to the following scaling of the interaction measure at high temperature:
\be
 \frac{e-3p}{T^4} \sim T^{-3\gamma -1}
\ee{IMSCALING}
Lattice QCD calculation results definitely favor $\gamma=1/3$.

\section{Power-law tailed distribution from repeated composition rules}

Power-law tailed distributions are numerous in physical phenomena, they can be viewed
as stationary distributions under certain constraints. A general approach in the framework
of non-extensive thermodynamics has been presented in Ref.\cite{BiroEPL2008}.
Starting with a general non-additive composition rule, 
its manifold repetition becomes asymptotically
an associative rule. Among these statistical attractors the rule
$h(x,y)=x+y+axy$ is particularly simple (here the simple addition is reconstructed
for $a=0$). It is asymptotic to any rule of $h(x,y)=x+y+G(xy)$ type.
We have shown that extreme relativistic kinematics can lead to such a rule for
kinetic energy composition if the pair interaction is a function of the relativistic
momentum transfer squared, $Q^2$, only. We surmise that parton cascades based on
pQCD cross sections in the large collision number limit may belong to this class
of effective composition rules.

Any associative composition rule can be expressed by the addition of
formal logarithms. In our case $(1+ah)=(1+ax)(1+ay)$ so the formal logarithm
is given by $L_a(x) = (1/a) \ln(1+ax)$.
The canonical energy distribution by this energy composition rule is given by
\be
 f_{{\rm eq}}(E) = e^{-\frac{L(E)-\mu}{T} } = e^{\mu/T} \, \left(1+aE \right)^{-1/aT}.
\ee{CANFORMAL}
This power-law tailed energy distribution is the Pareto-Tsallis distribution
with $q=1+aT$.
Its reciprocal logarithmic slope is proportional to the energy:
\be
 {\cal T}(E) = T + (q-1)E.
\ee{TSLOPE}

\section{Coalescence quark scaling in the power-law}

Besides the familiar factorizing coalescence formula valid at high energies and momenta,
we consider a hadronization kinematics with string like objects contributing
to the total energy of hadrons. Regarding an energy balance of
\be
 E_1 + E_2 + \sigma \ell = \sqrt{m_H^2+(\vec{p}_1+\vec{p}_2)^2},
\ee{STRENERG}
for meson formation (and a three-fold one for baryon formation) we consider
\ba
 f(E_H) &=& \int\!\! d^3p dE_1 d^3q dE_2 dm \, f_1\left(\frac{1}{2}\vec{p}+\vec{q}\right) 
		f_2\left(\frac{1}{2}\vec{p}-\vec{q}\right) \nl 
 & & C(m,q) \, g(m) \, \Theta\left( E_H^2 - (E_1+E_2+m)^2 \right)
\ea{COALES}
for the coalesced hadron spectra. Here $\Theta(x)$ denotes the step function,
and the integration variable $m=\sigma \ell$ runs over the energy values put into a string.
$C(q,m)$ is the coalescence factor, about which here we assume that prefers very
low relative momenta only and put $C(q,m)= c \, \delta^3(\vec{q})$. 
About the string length distribution we assume the form
\be
 g(m) \sim m^{d-1} \exp \left(-\left[\Gamma(1+1/d) \frac{m}{\langle m \rangle} \right]^d \right)
\ee{STRDIST}
argued for in Ref.\cite{BiroShan2003}.
Here the effective dimensionality of the moving string, $d$, and the average energy
stored in it, $\langle m \rangle$ are parameters. The coalescence ratio can then be
interpreted in terms of the string mass distribution:
\be
\frac{f_H(E_H)}{f_1(E_1)f_2(E_2)} = \int_{E_H-E_1-E_2}^{\infty} g(m)dm.
\ee{FACTOR}

\begin{figure}
\begin{center}
\includegraphics[width=0.47\textwidth]{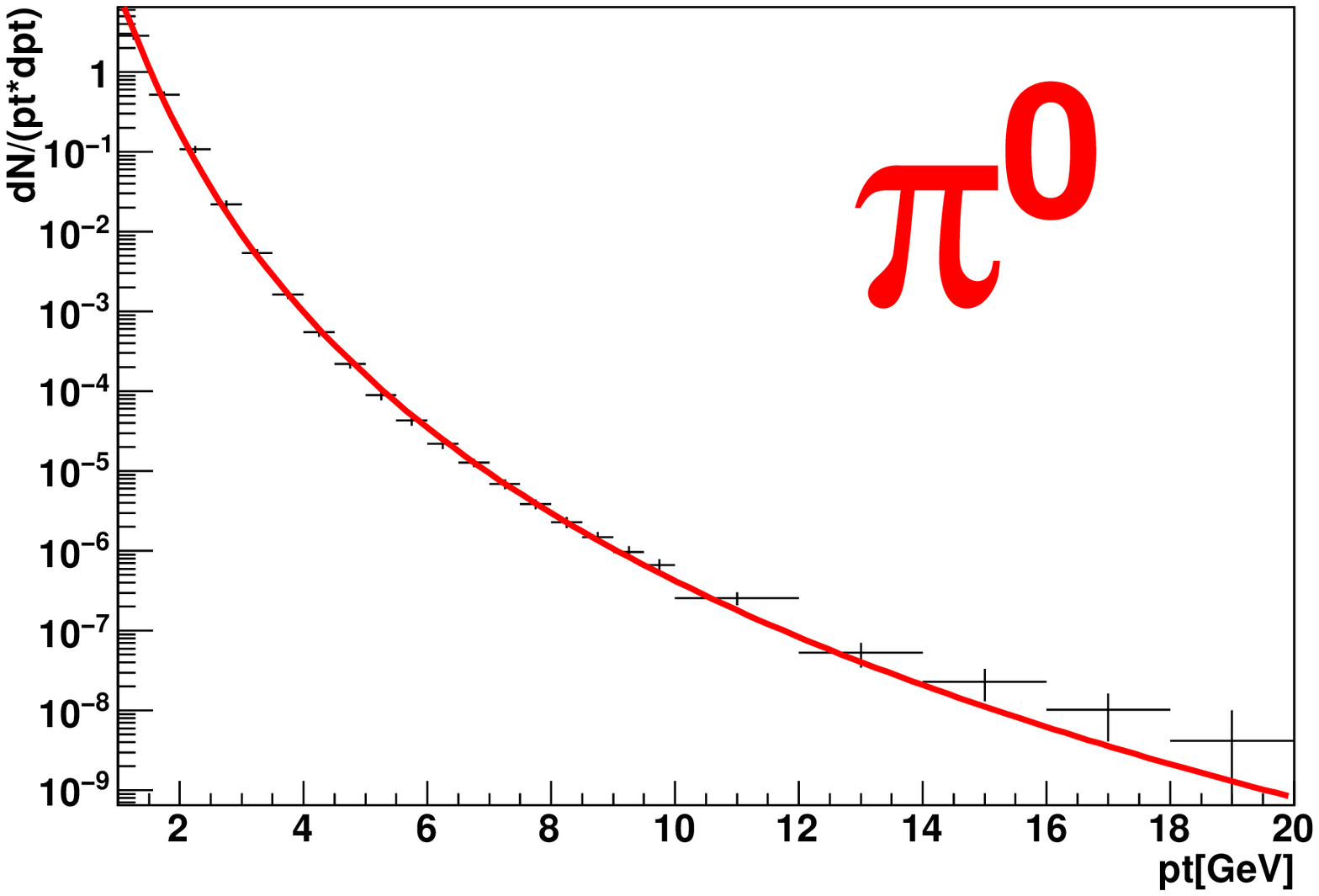}%
\includegraphics[width=0.47\textwidth]{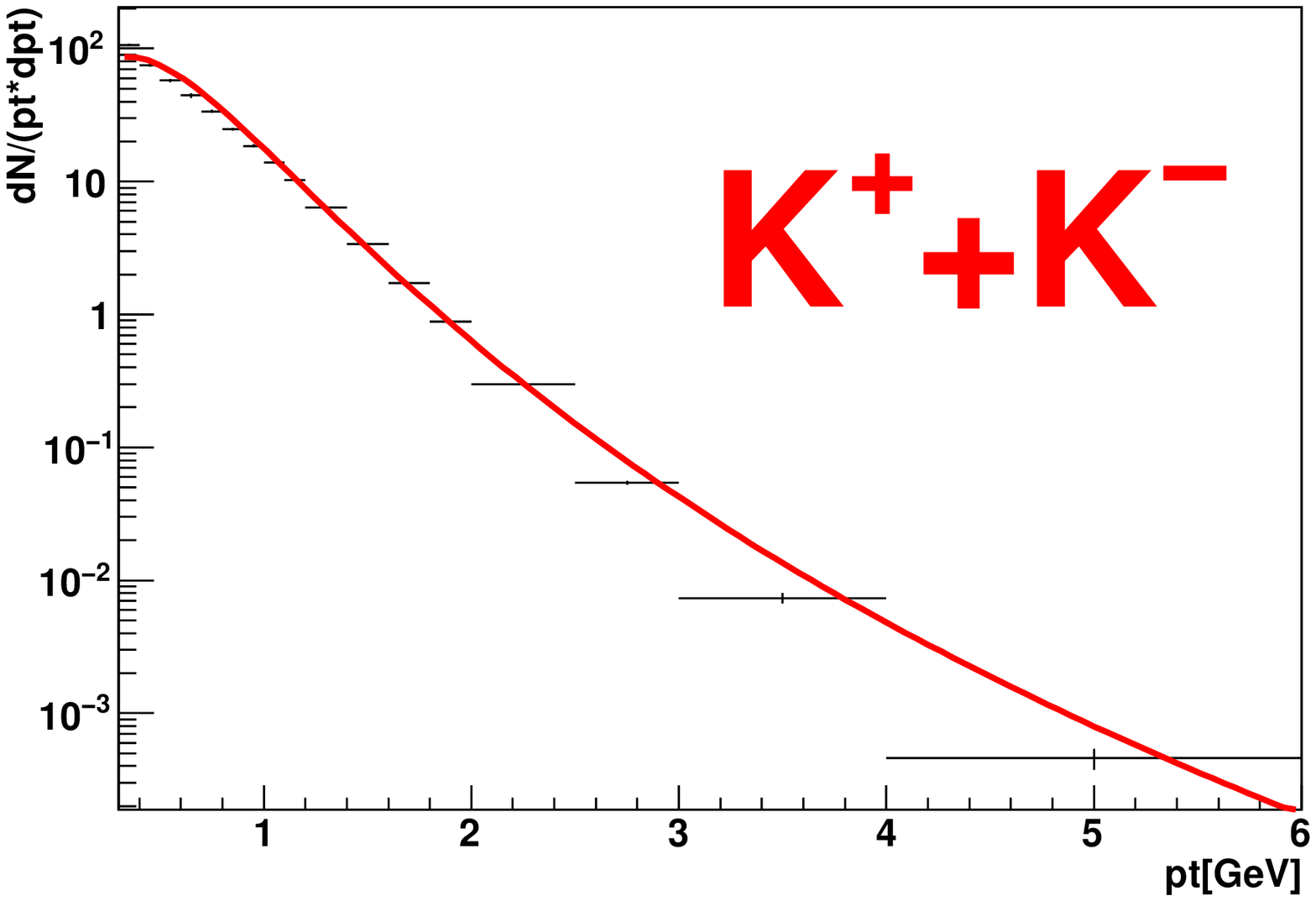}
\includegraphics[width=0.47\textwidth]{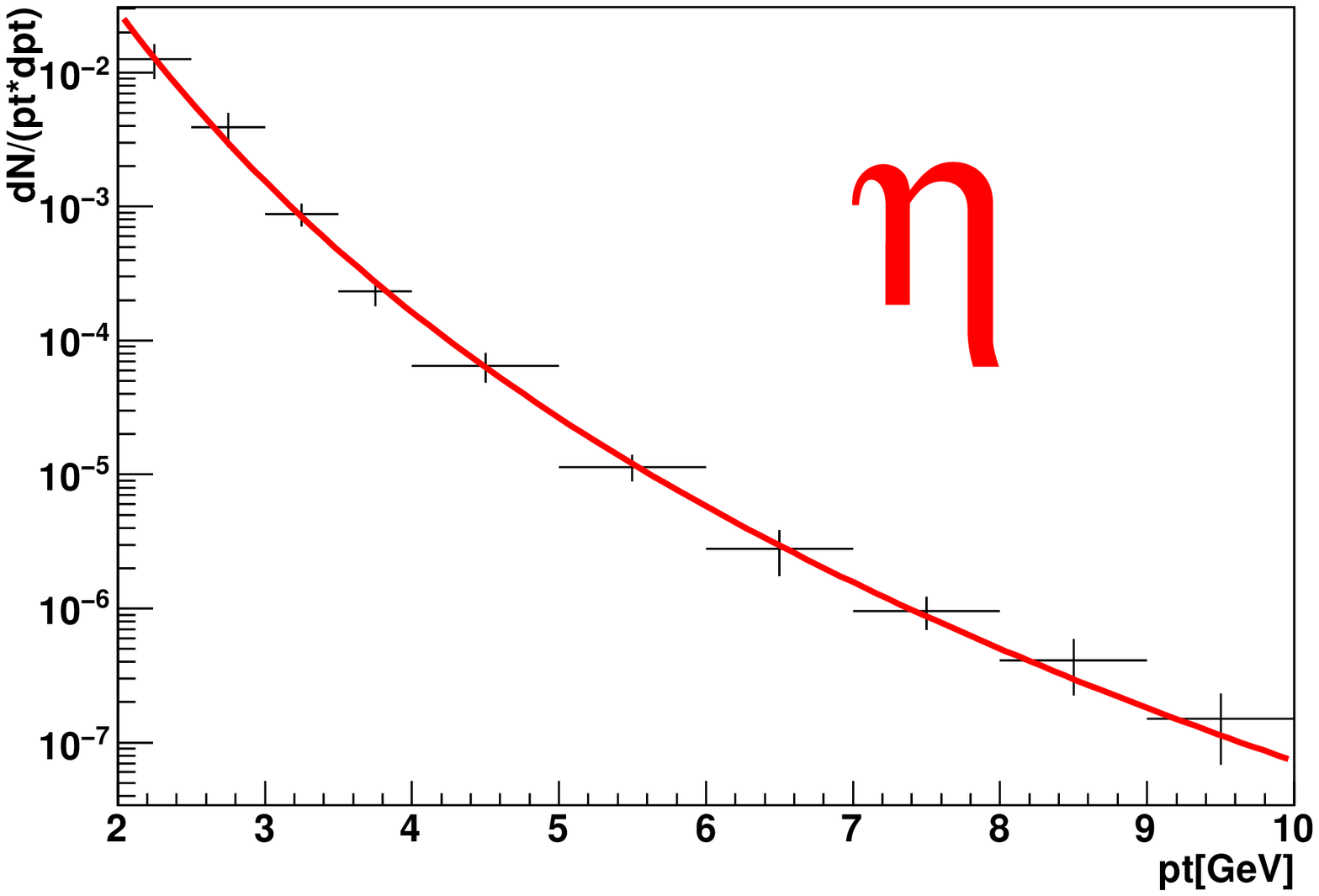}%
\includegraphics[width=0.47\textwidth]{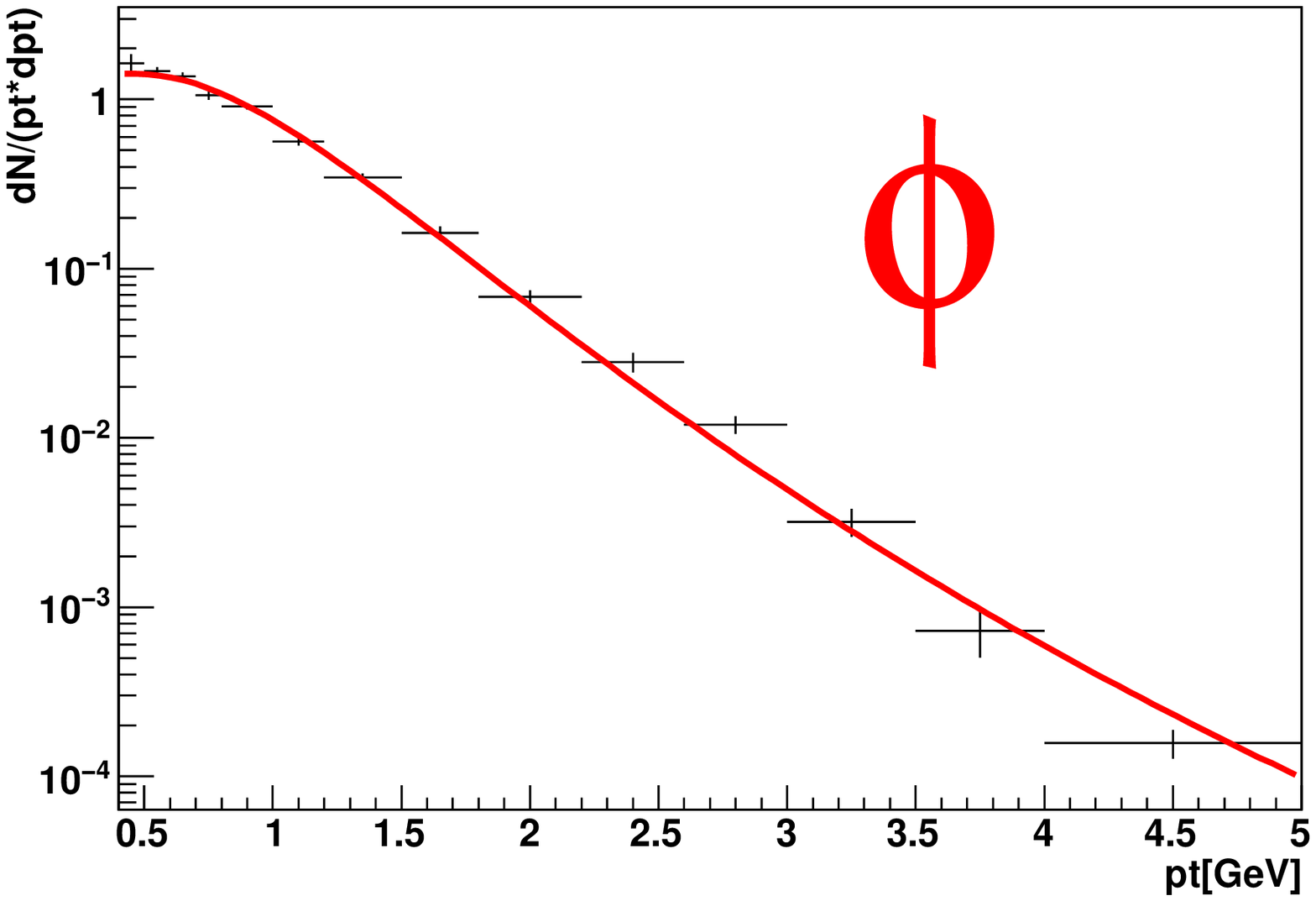}
\includegraphics[width=0.47\textwidth]{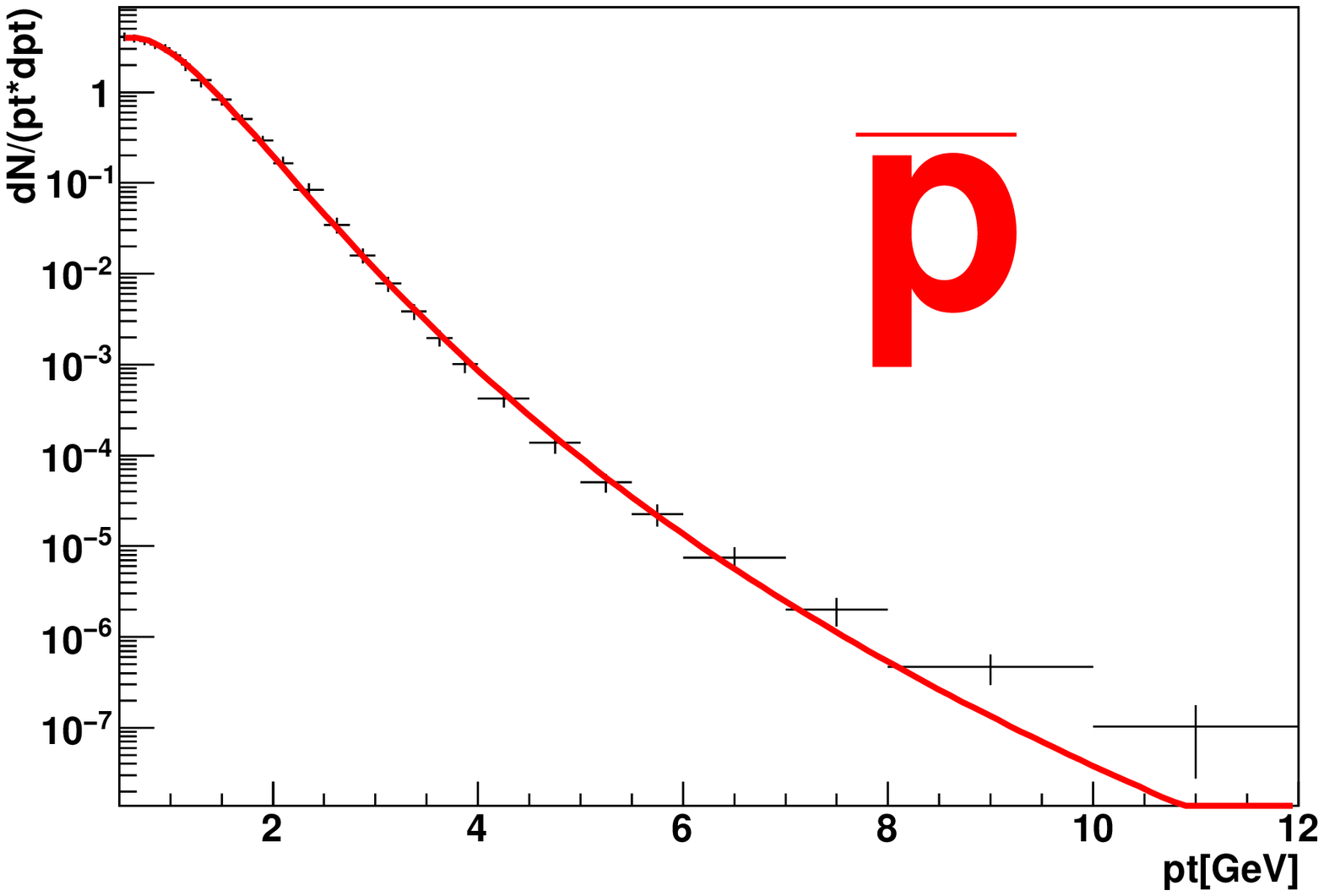}%
\includegraphics[width=0.47\textwidth]{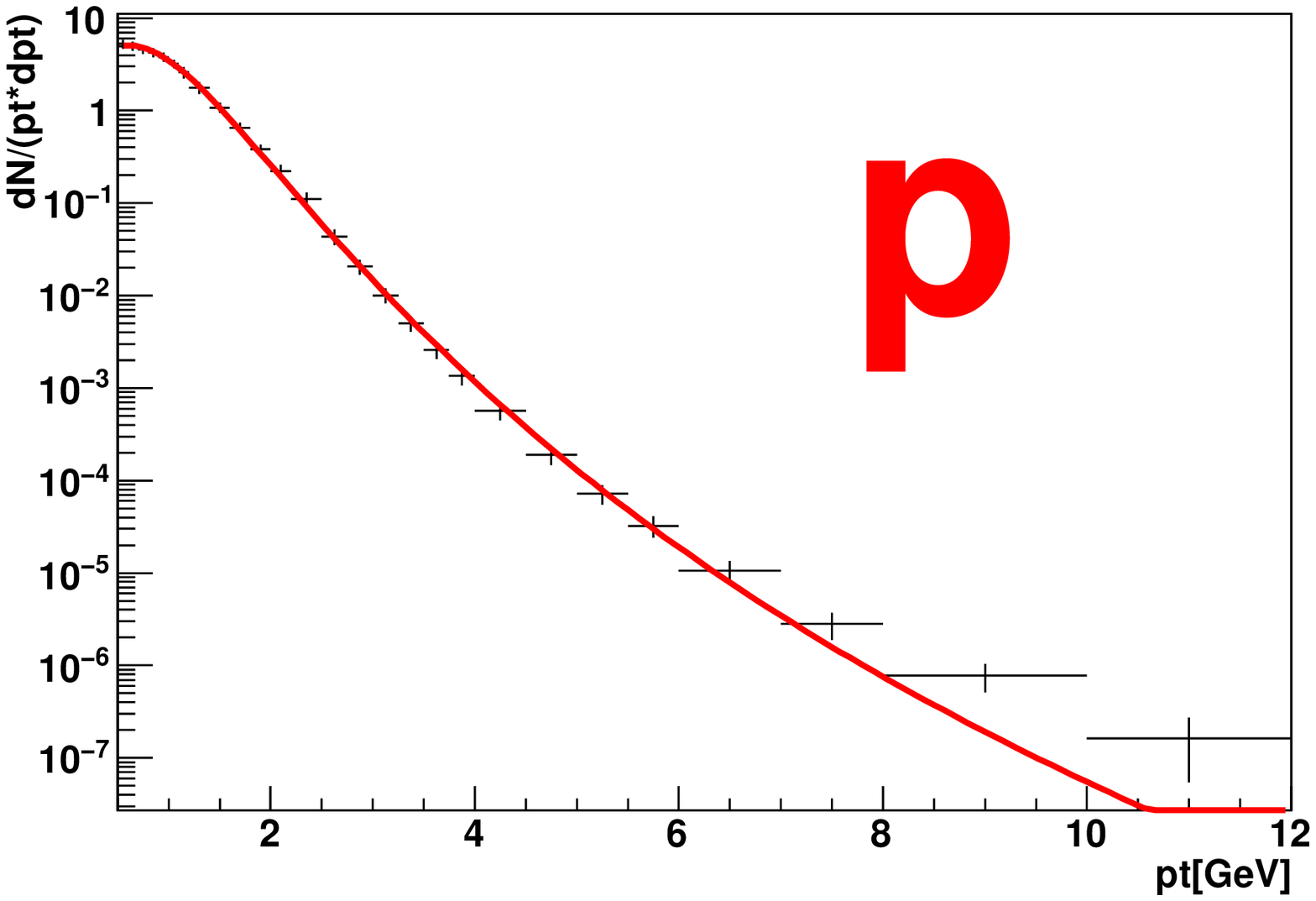}
\includegraphics[width=0.47\textwidth]{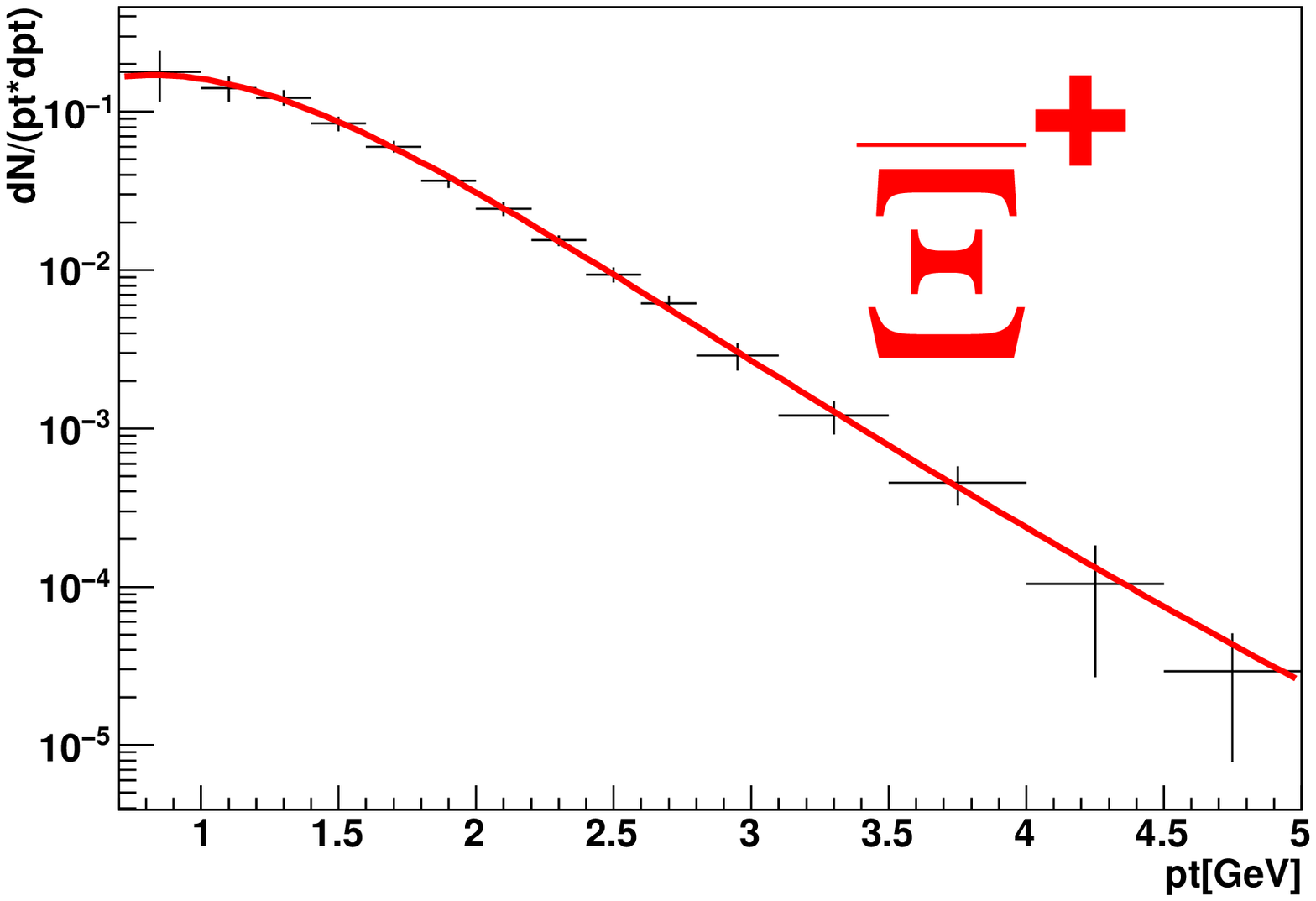}%
\includegraphics[width=0.47\textwidth]{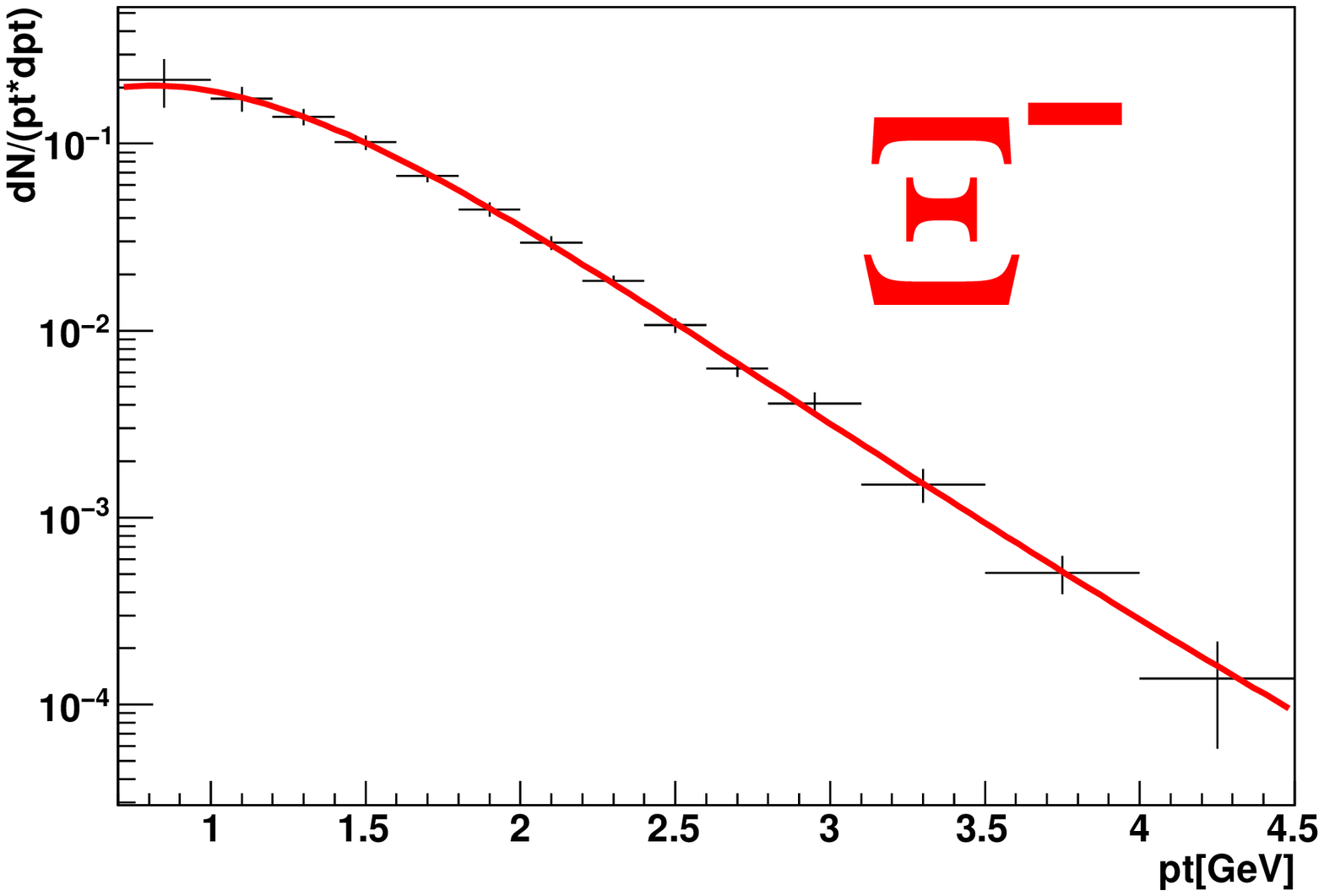}
\includegraphics[width=0.47\textwidth]{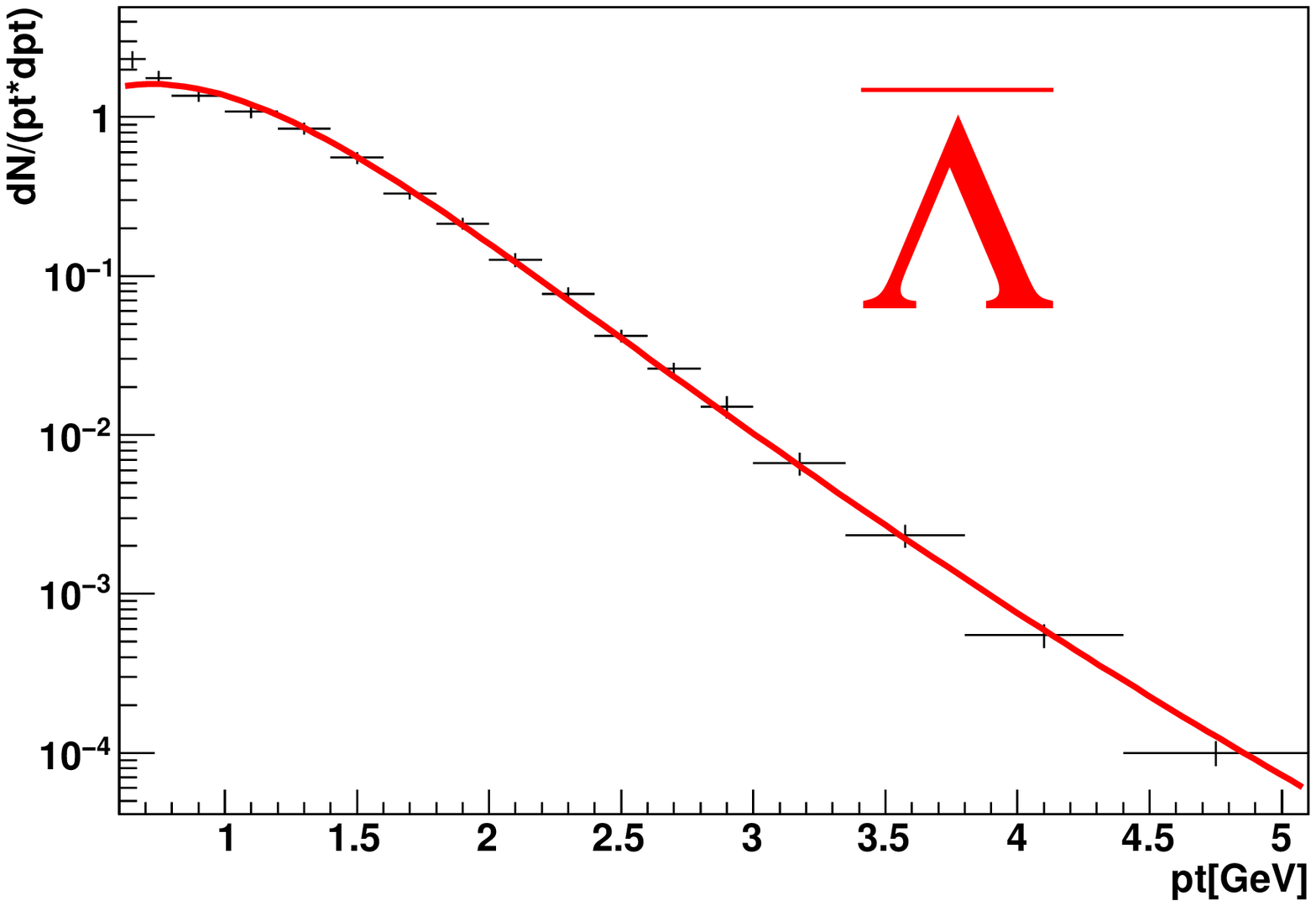}%
\includegraphics[width=0.47\textwidth]{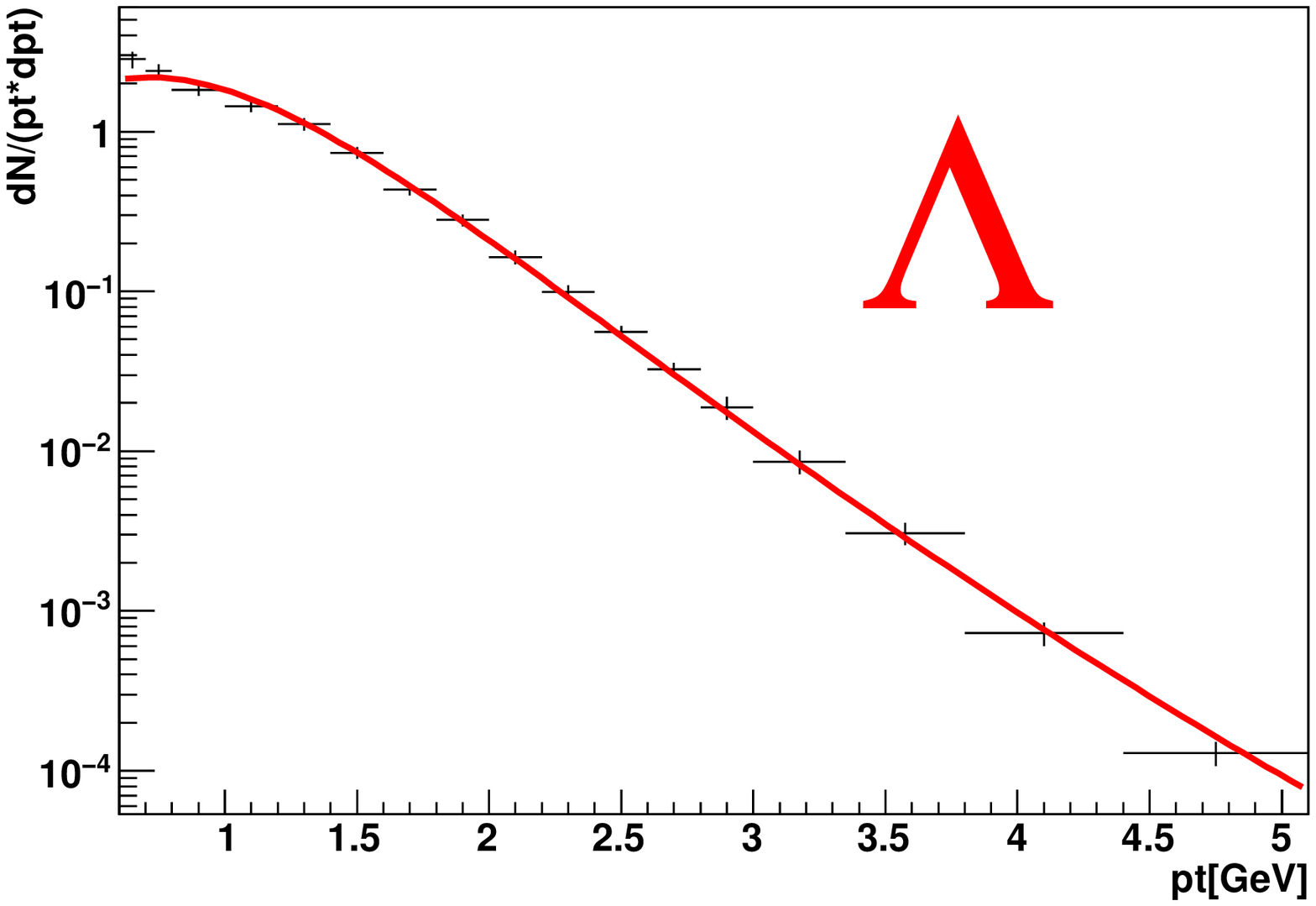}
\end{center}
\caption{\label{MESONFIT}
  Tsallis distribution fit to different hadronic transverse momentum spectra 
  with a transverse blast wave taken into account.
}
\end{figure}

\vspace{3mm}
\section{Results}

In the following figures thermal Tsallis-Pareto distributions are fitted to
several hadronic transverse momentum spectra obtained at RHIC by the PHENIX\cite{PHENIX1,PHENIX2}
and STAR\cite{STAR1,STAR2,STAR3,STAR4} groups. 
In Fig.\ref{MESONFIT} the differential yields for pions, kaons,
eta and phi mesons, protons, Lambda-s and Xi hyperons are shown including their antiparticle
partners.
The red continuous lines represent Tsallis-Pareto fits with best parameters 
for each hadron. Transforming these spectra in a frame co-moving with a transverse
blast wave, the co-moving energy distribution $f(E)$ is given in terms of $E=\gamma(m_T-vp_T)$.
Since this transformation is mass dependent, the $p_T$ spectra show maxima at different
values $p_T^{{\rm max}}=\gamma v m$.

\begin{figure}
\begin{center}
\includegraphics[width=0.95\textwidth]{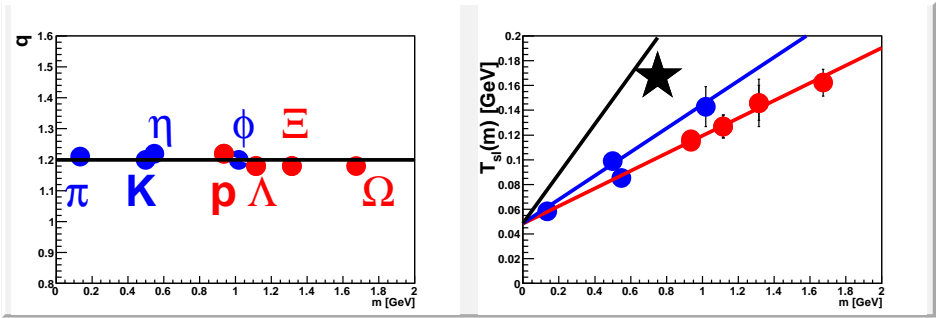}
\caption{\label{FITPARAMS}
  Tsallis distribution fit  parameters for quark matter extracted from hadron spectra.
  The slope scales linearly with the mass due the Tsallis formula and follows
  the valence quark number scaling indicated by the red and blue lines.
  The black line belongs to the conjectured Tsallis-Pareto distributed quark matter,
  the star denotes the average hadron mass value of $750$ MeV extracted from
  $E/N=m+3T/2=1$ GeV at $T_c=175$ MeV temperature. 
}
\end{center}
\end{figure}

We assume that already the quark matter had this collective transverse flow at
the hadronization, only the hadron spectra were formed according to 
the coalescence formula eq.(\ref{FACTOR}). Fig.\ref{FITPARAMS} plots
the extracted quark matter $q$-value from different 
hadron spectra (upper part). This value is quite stable around $q \approx 1.2$.
The slope parameters (lower part) at the minimal co-moving energy, $E=m$,
(meaning $p_T=\gamma v m$, the position of the maxima) are proportional
to the hadron mass due to eq.(\ref{TSLOPE}). This proportionality
can be observed even after disentangling the transverse flow effect, in contrast
to non-Tsallis ($q=1$) fits done earlier to heavy ion data.

\begin{figure}
\begin{center}
\noindent\hfill\includegraphics[scale=0.16]{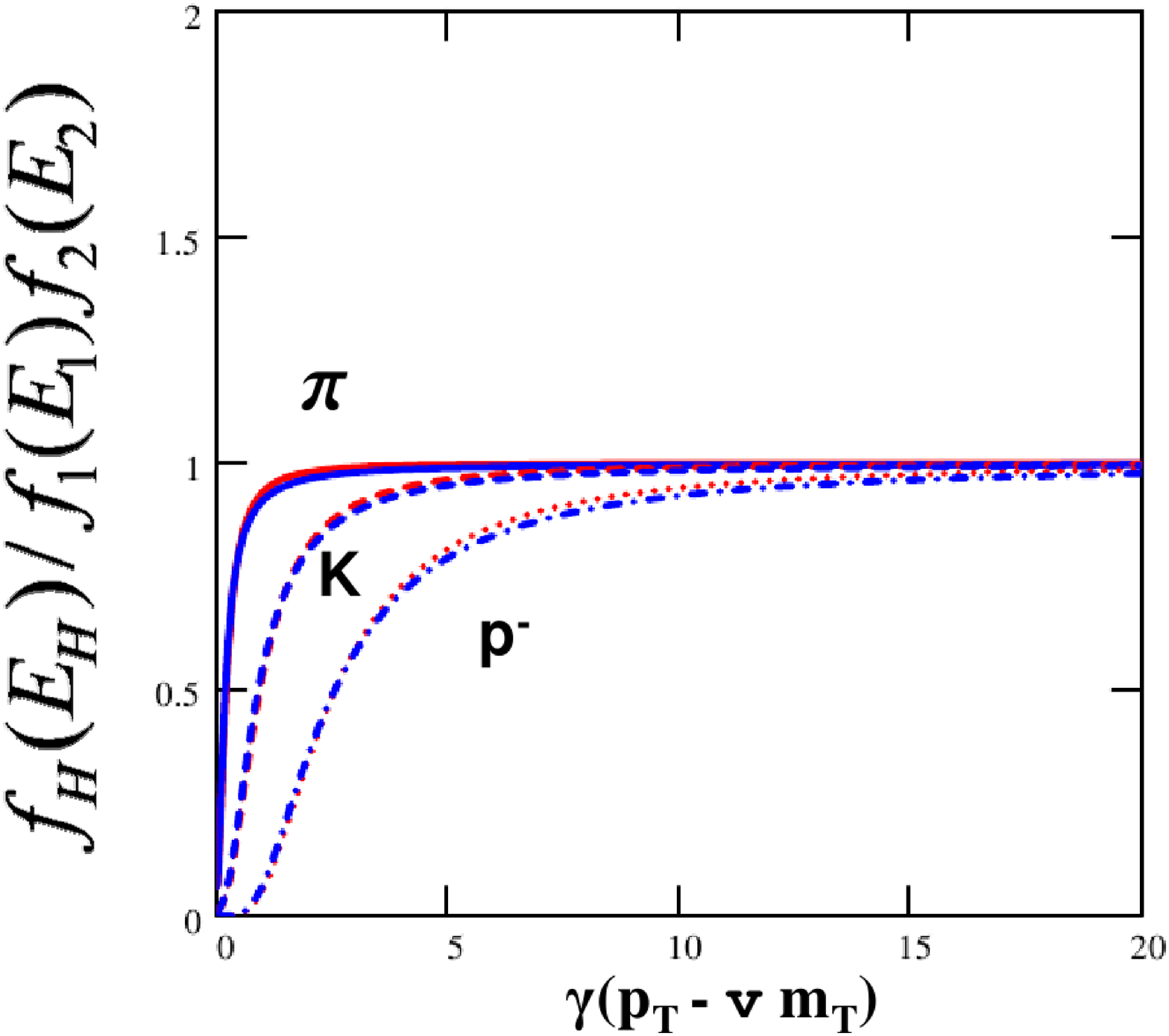}\ \ %
\includegraphics[scale=0.16]{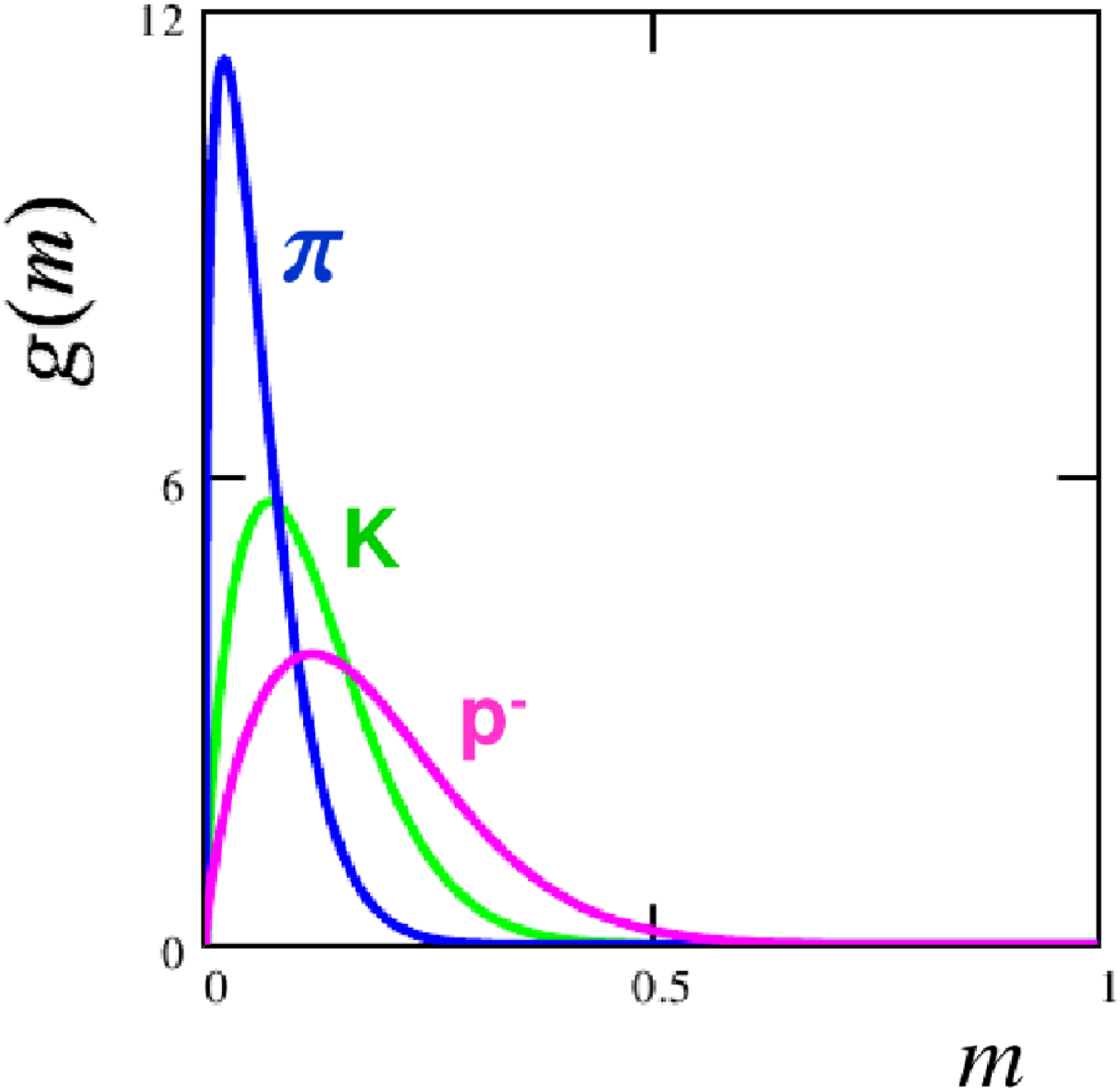}
\end{center}
\caption{\label{STRINGS}
    Violation of the coalescence product rule at low energy (left)
    Prediction for the differential string-energy distributions by coalescence from
    Tsallis-Pareto distributed quark matter and strings (right).
}
\end{figure}

The factorization of hadron spectra to the respective quark spectra,
as predicted by the naive coalescence model, is well satisfied at high
momenta. For lower momenta, however, this scaling is violated.
It can be seen by the ratio $f_H/f_1f_2$ shown in Fig.\ref{STRINGS} (left part).

In order to explain this violation we supplement the quark coalescence picture
by that of strings built in into the forming hadrons.
Assuming that all strings longer than a threshold value contribute with 
uniform probability to the corresponding hadron yield, 
the missing energy in the coalescence
model can be attributed to strings, distributed in their length.
Fig.\ref{STRINGS} (right part) plots the differential string energy distribution 
$g(m)$.
Based on experimental pion, kaon and antiproton spectra in RHIC STAR and PHENIX experiments,
we fit the average mass contribution values $\langle m \rangle_{\pi} = 60$ MeV,
$\langle m \rangle_{K} = 120$ MeV, $\langle m \rangle_{\overline{p}} = 185$ MeV
 and dimensionalities $d_{\pi}=1.30$, $d_{K}=1.60$ and $d_{\overline{p}}=1.67$.

\section*{Acknowledgment}

This work has been supported by the Hungarian National Research Fund OTKA
and the National Bureau for Research and Development (K49466 and K68108).

\end{document}